\documentclass[12pt]{iopart}
\usepackage{graphicx,latexsym}
\usepackage[square,sort&compress,numbers]{natbib}
\usepackage{pifont}
\newcommand{\cmark}{\ding{51}}%
\newcommand{\xmark}{\ding{55}}

\newcommand{\be}{\begin{equation}}
\newcommand{\ee}{\end{equation}}
\newcommand{\ba}{\begin{eqnarray}}
\newcommand{\ea}{\end{eqnarray}}

\begin{document}

\title[Cluster formation and phase separation in HJD]%
{Cluster formation and phase separation in
heteronuclear Janus dumbbells}

\author{G Muna\`o$^1$%
\footnote{Corresponding author: gmunao@unime.it},
P O'Toole$^2$, T S Hudson$^2$, D Costa$^1$, \\ C Caccamo$^1$,
F Sciortino$^3$ and  A Giacometti$^4$ }
\address{
$^1$Dipartimento di Fisica e di Scienze della Terra,
Universit\`a degli Studi di Messina,
Viale F.~Stagno d'Alcontres 31, 98166 Messina, Italy}
\address{
$^2$School of Chemistry, University of Sidney,
NSW 2006, Australia}
\address{
$^3$Dipartimento di Fisica,
Universit\`a di Roma ``Sapienza'',
Piazzale Aldo Moro 2, 00185 Roma, Italy}
\address{
$^4$Dipartimento di Scienze Molecolari e Nanosistemi,
Universit\`a Ca' Foscari Venezia,
Calle Larga S.~Marta DD2137, Venezia I-30123, Italy}

\begin{abstract}
We have recently investigated the phase behaviour of model
colloidal dumbbells constituted by two identical tangent hard spheres,
with the first one being surrounded by an attractive square-well interaction
(Janus dumbbells,
Muna\'o G {\it et al} 2014 {\it Soft Matter} {\bf 10} 5269).
Here we extend our previous analysis by introducing in the model
the size asymmetry
of the hard-core diameters, and study the
enriched phase scenario thereby obtained.
By employing standard Monte Carlo simulations we show that in such
``heteronuclear Janus dumbbells''
a larger hard-sphere site promotes the formation of clusters, whereas
in the opposite condition a gas-liquid phase separation takes place,
with a narrow interval of intermediate asymmetries wherein the two phase
behaviours may compete.
In addition, some peculiar geometrical arrangements, such as lamell\ae,
are observed only around the perfectly symmetric case.
A qualitative agreement is found with
recent experimental results, where it is shown that
the roughness of molecular surfaces in heterogeneous dimers leads to 
the formation of colloidal micelles.

\end{abstract}

\section{Introduction}
Colloidal dumbbells are currently object of rather intense
experimental~%
\cite{Nagao:11,Fossum:14,Riley:14,Yoon-Chem,Forster,Ma:12,Hosein}
and theoretical investigations~%
\cite{Fejer:14,Munao:PCCP,Munao:14},
due to the possibility offered by such particles
to act as building blocks for the fabrication
of new materials~\cite{Yoon-Chem},  such as photonic crystals~\cite{Forster},
self-assembled structures under the effect of electric fields~\cite{Ma:12}
and other complex structures~\cite{Hosein}.
One key feature of such dumbbells is the
asymmetry in the relative size of the constituting spheres and/or in their
interaction potential.
Pure hard-sphere, as well as pure square-well colloidal
dumbbells have been widely studied, with a
variety of investigation concerning
their  thermodynamic and structural properties~\cite{Dijkstra:13,Munao-cpl,%
Munao:09,Dijkstra-pre,Dijkstra-jcp,Lowen:11,Dijkstra:12,Chapela:11,Miller:09,Del-gado:11,Chong-prl}.
In the special case in which
one of the two particles
is solvophilic, and the other one is solvophobic,
the ``molecule'' represents a simple example of
a colloidal surfactant (Janus dumbbell~\cite{Janusdumbbell,Janusdumbbell1}).
Such systems  constitute a
molecular
generalization of the well-known concept of
Janus spherical particles~%
\cite{Stellacci,Janus,Janusgold,%
granick,Janusweitz,Adv:10,Chen:11}, largely
investigated because of the rich variety of self-assembled
structures they may form. In the Janus dumbbell case, such a scenario may be
further enriched by the possibility to tune the aspect ratio and the
asymmetry of the dumbbell, thus originating what we shall henceforth term
heteronuclear
Janus dumbbells (HJD). It has been recently shown that HJD,
under appropriate conditions, are able to self-assemble in
colloidal micelles~\cite{Kraft:12}, promoted by the surface roughness of the
particles; they may
form colloidal molecules as well as larger supracolloidal structures~\cite{Bon:14}.
In spite of the scientific and technological importance of HJD,
few simulation studies
have focused on such systems: in particular, an assessment of experimental results
in comparison with simulation data has been carried out only
in~\cite{Kraft:12,Bon:14}.
Therefore, more investigations concerning
the phase behaviour of HJD would be highly desirable, in particular to study
how the heterogeneity influences the competition between the formation
of aggregates and phase separation.
From a microscopic viewpoint,
such a competition has been generally interpreted
in terms of simple spherical models
characterized by
the simultaneous presence of short-range attractions and long-range repulsions
in the total interaction
(see e.g.~\cite{Cardinaux:07,Costa:10,Bomont:12,Godfrin:13, Godfrin:14}
and references).
Physically, the long-range repulsion generally stems from 
the weakly screened charge carried by colloidal macromolecules,
whereas the short-range attraction stems
from several different mechanisms,
including depletion forces, van der Waals interactions,
hydrophobic effects~\cite{Saville}.
In this picture,
the formation of clusters out of the homogeneous fluid
is intepreted as due to an appropriate balance
between attraction, promoting
the formation of aggegates at low temperature, 
and long-range repulsion, preventing a complete phase
separation~\cite{Pini:00,Imperio:04}.
More recently, 
the interplay between the aggregate formation
and phase separation
has been investigated by means of various theoretical and simulation tools
also for more sophisticated models, as for instance 
patchy~\cite{Doye:11,prl-lisbona,Lorenzo:14} 
and Janus~\cite{janusprl,pccpjanus} particles.

In this work 
HJD are modelled as two tangent hard spheres, with the first one
being surrounded by a square-well interaction with fixed attraction range;
the heterogeneity is introduced 
by changing the ratio between the two hard-core diameters.
We show by standard Monte Carlo simulations
that even moderate asymmetries sensitively influence the overall phase scenario,
giving rise to
a competition between the gas-liquid phase separation and
the spontaneous formation of different self-assembled structures in the fluid.
We compare with our previous investigation of
homonuclear Janus dumbbells~\cite{Munao:14}
and show that the development of peculiar planar structures (lamell\ae)
therein  observed at low temperatures,
is only found around the perfectly symmetric case.
At variance,  the gas-liquid
phase separation, turning to be completely suppressed
in the homonuclear case~\cite{Munao:14}, reveals again
if the sphere  bearing the square-well interaction becomes larger
than the second one.

The paper is organized as follows:
in Section~2 we detail the interaction properties of the HJD model
and the simulation technique we have employed in this study. Results
are reported and discussed in Section~3. Conclusions follow in Section~4.

\section{Models and methods}

\begin{figure}[!h]
\begin{center}
\includegraphics[width=8.0cm,angle=0]{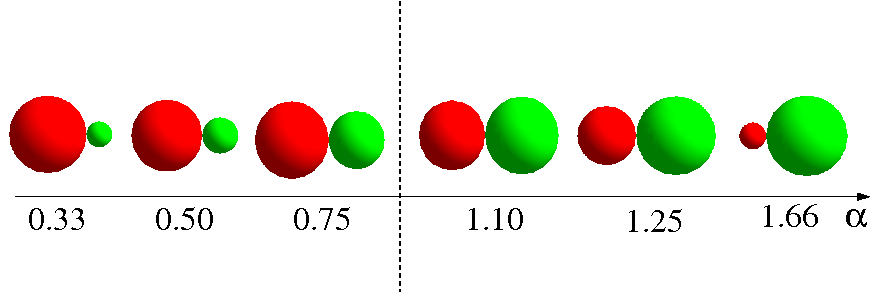}
\caption{Schematic representation of HJD for
all $\alpha$ values investigated in this work, see equation~(\ref{eq:alpha}).
In each pair, left (red) and right (green) spheres correspond
to the HS and SW sites, respectively. }
\label{fig:models}
\end{center}
\end{figure}

Our dumbbell model is constituted by two tangent hard spheres~---~%
characterized by different core diameters $\sigma_1$ and $\sigma_2$~---~%
with the first one being surrounded by a square-well attraction;
the interaction potential among sites $i$ and $j$ ($i, j \in [1,2]$)
on different molecules is then written as:
\numparts
\ba
\label{eq:pot1}
U_{11}(r)=\cases{
\infty & if $r < \sigma_1$ \\
-\varepsilon & if $\sigma_1 \le r < \sigma_1+\lambda\sigma_1$ \\
0 & otherwise} \\
\label{eq:pot2}
U_{12}(r) = U_{21}(r)=\cases{
\infty & if $r < (\sigma_1+\sigma_2)/2$ \\
0 & otherwise} \\
\label{eq:pot3}
U_{22}(r)  = \cases{
\infty & $r < \sigma_2$ \\
0 & otherwise}
\ea
\endnumparts
The
diameters $\sigma_1$ and $\sigma_2$
are defined in terms of a unit of length $\sigma$ and a parameter $\alpha$,
so that:
\ba\label{eq:alpha}
\cases{\sigma_1=\alpha\sigma\\
\sigma_2=\sigma}
 \qquad {\rm if} \ 0 < \alpha\leq 1\,; \qquad
\cases{
\sigma_1=\sigma \\
\sigma_2=(2-\alpha)\sigma}
 \qquad {\rm if} \ 1 < \alpha\leq 2
\ea
The two limits $\alpha=0$ and $\alpha=2$ correspond
to pure hard-sphere (HS) and pure square-well (SW)
atomic fluids respectively, whereas $\alpha=1$
is the symmetric case.
In the $\alpha=0$ limit, neither a gas-liquid
transition nor the formation of clusters take place whereas, in the
$\alpha=2$ limit,
a conventional gas-liquid phase separation (whose critical parameters depend on
the width of the attractive square-well) occurs and
no cluster formation is expected. Finally,
for $\alpha=1$ we recover the homonuclear
Janus dumbbell configuration for which
(in the case $\lambda=0.5$)
the gas-liquid phase separation is absent and both spherical
and planar (lamell\ae) clusters are observed~\cite{Munao:14}.
Models investigated in this work correspond to the size parameter $\alpha$
varying in the range $0.33 \leq \alpha \leq 1.66$ and are schematically
depicted in figure~\ref{fig:models}:
the evolution with increasing $\alpha$ from a HS dominating (in size)
configuration to the opposite SW dominating one, is therein illustrated.
The well depth $\varepsilon$ in equation~(\ref{eq:pot1}) gives
the unit of energy, in terms of which
we define the reduced temperature $T^*=k_{\rm B}T/\varepsilon$
(with $k_{\rm B}$ as the Boltzmann constant);
we also introduce the reduced density $\rho^*=(N/V)\sigma^3$
(where $N$ is the number of particles
and $V$ the volume).
In all calculations we also set, as in ~\cite{Munao:14},  $\lambda=0.5$ 
in the definition~(\ref{eq:pot1})
of $U_{11}(r)$.

In order to characterize self-assembled structures and thermodynamic
properties of HJD, we have carried out standard Monte Carlo simulations of
a sample composed by 500 particles enclosed in a cubic box with periodic boundary conditions
at four different densities (specifically, $\rho^*=0.05$,
$\rho^*=0.10$, $\rho^*=0.20$ and $\rho^*=0.30$) and several temperatures in
the whole investigated range of $\alpha$-values.

We
shall make use of the second virial coefficient
$B_2$, written for a molecular system as~\cite{Hansennew}:
\begin{equation}\label{eq:b2int}
B_2(\beta\epsilon)=-\frac{1}{2}\int f_{ij}({\bf r}) d {\bf r} d\mathbf \Omega_i
d\mathbf \Omega_j
\end{equation}
where $\beta=1/T^*$,
$f_{ij}({\bf r})=\exp[-\beta U_{ij}({\bf r})]-1$
is the Mayer function between molecules $i$ and $j$ and
$U_{ij}({\bf r})$ is
the intermolecular potential defined by equations~(\ref{eq:pot1})-(\ref{eq:pot3}).
Moreover, in equation~(\ref{eq:b2int})
$\int \dots d\mathbf \Omega$ represents
the integration over all the orientations, normalized so that
$\int d\mathbf \Omega = 1$.
Following the method employed by Yethiraj and Hall~\cite{Yeth-molphys}, we have
numerically computed $B_2$
by generating a large number $N_{\rm c}$ of independent configurations of
two dumbbells in a cubic box of side $L$; then,  by averaging the
Mayer function over
all such configurations, we obtain:
\begin{equation}\label{eq:b2}
B_2(\beta\epsilon)=-\frac{L^3\langle f_{ij} \rangle}{2N_c}\,.
\end{equation}
Values of $B_2 < 0$ ($B_2 > 0$) indicate that attractive (repulsive)
interactions are prevailing, with
the Boyle temperature
$T_{\rm B}^*$, at which $B_2=0$, separating the two regimes.
In the next section we shall employ $T_{\rm B}^*$
to locate the regions in the $\alpha-T^*$ plane over which the
attractive interactions are more effective.

\begin{figure}[!h]
\begin{center}
\includegraphics[width=8.0cm,angle=0]{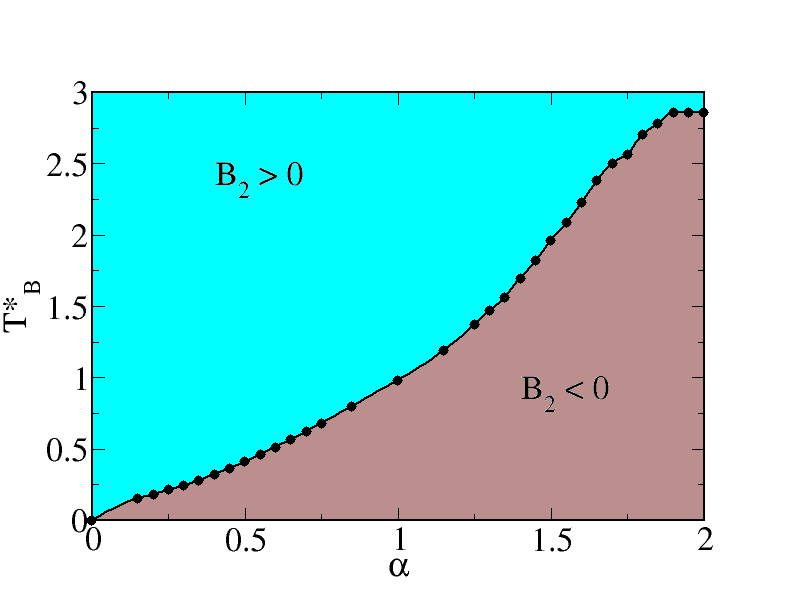}
\caption{Boyle temperature $T_{\rm B}^*$ (circles, with the line as a guide to the eye)
vs $\alpha$. The upper (cyan) and lower (brown) regions
correspond, respectively, to positive and negative values of
the second virial coefficient $B_2$.}
\label{fig:Boyle}
\end{center}
\end{figure}

\section{Results and discussion}

Using equation~(\ref{eq:b2}), we first determine $T^*_{\rm B}$ as explained in
the previous Section.
Results as a function of $\alpha$ are reported in figure~\ref{fig:Boyle}:
as visible,
the $T_{\rm B}^*$ vs $\alpha$ locus delimits two regions,
over which $B_2 > 0$ and $B_2 < 0$, respectively.
In the $B_2 > 0$ region,
HJD should not experience enough attractions to give rise to significant
particle association.
Conversely, in the $B_2 < 0$ region,
attractive interactions are effective and
clustering or droplet formation may occur;
it appears that
such a region becomes smaller if $\alpha$ decreases,
and disappears in the limit of $\alpha=0$.
In the other limit, in which $\alpha$ approaches 2, i.e. for purely SW
spheres,
$T^*_{\rm B}$ gets a saturation value over a limited range of $\alpha$, thus
suggesting the phase behaviour of the ordinary SW fluid to moderately
extend inside the $\alpha-T^*$ plane.

We now examine in more detail various
heterogeneity regimes.

\begin{figure}[t!]
\begin{center}
\begin{tabular}{lr}
\includegraphics[width=8.0cm,angle=0]{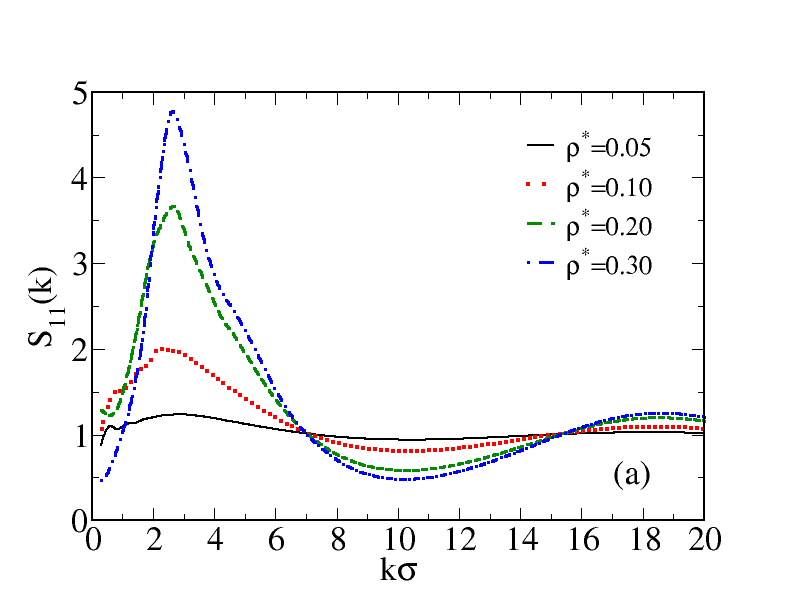} &
\includegraphics[width=8.0cm,angle=0]{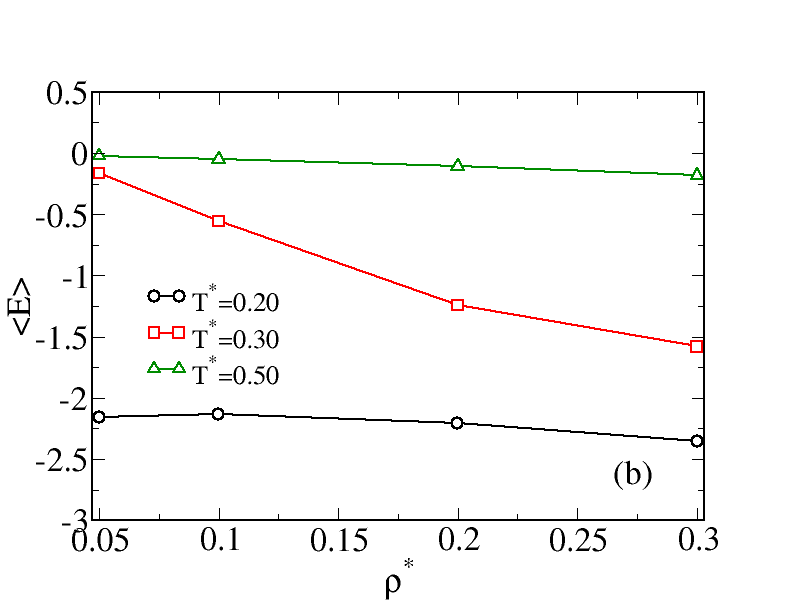}
\end{tabular}
\caption{Panel (a): $S_{11}(k)$
for $\alpha=0.33$ at $T^*=0.30$ and several densities.
Panel (b): average internal energy per particle $\langle E \rangle$
vs $\rho^*$ along different isotherms.}\label{fig:alfa-033}
\end{center}
\end{figure}

\subsection{Cluster formation (${0.33<\alpha<1}$)}\label{sec:3.1}

\begin{figure}[!t]
\begin{center}
\begin{tabular}{l}
\includegraphics[width=3.8cm,angle=0]{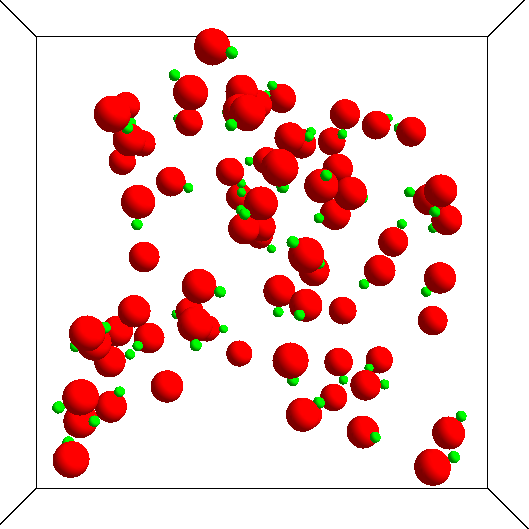}
\includegraphics[width=3.8cm,angle=0]{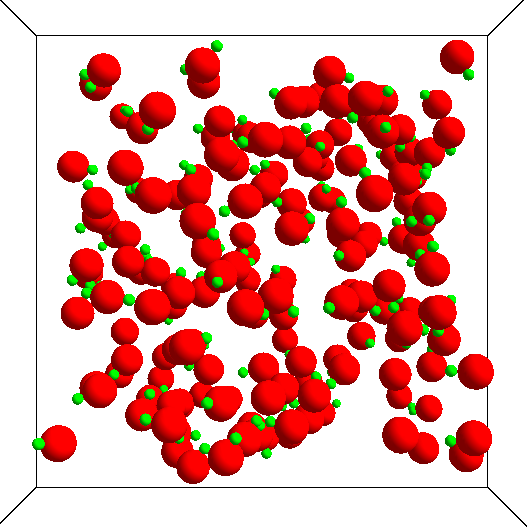}
\includegraphics[width=3.8cm,angle=0]{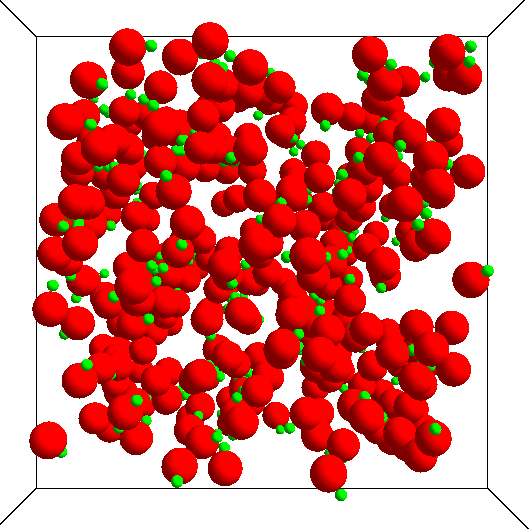}
\includegraphics[width=3.8cm,angle=0]{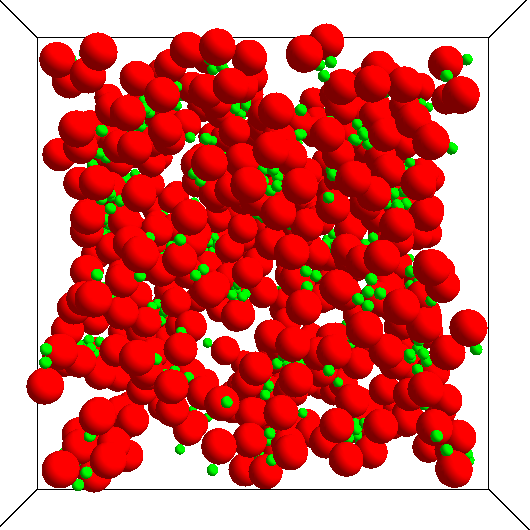} \\
\includegraphics[width=3.8cm,angle=0]{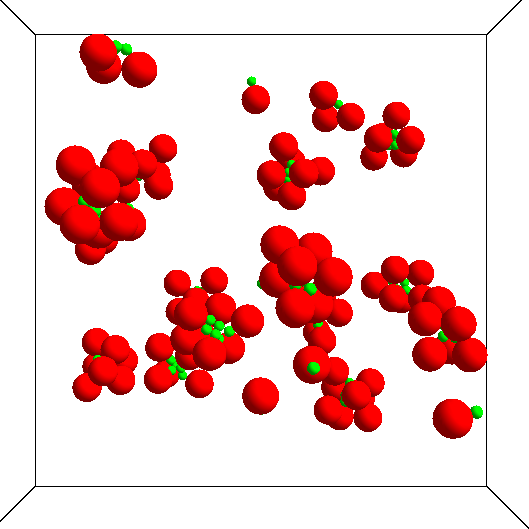}
\includegraphics[width=3.8cm,angle=0]{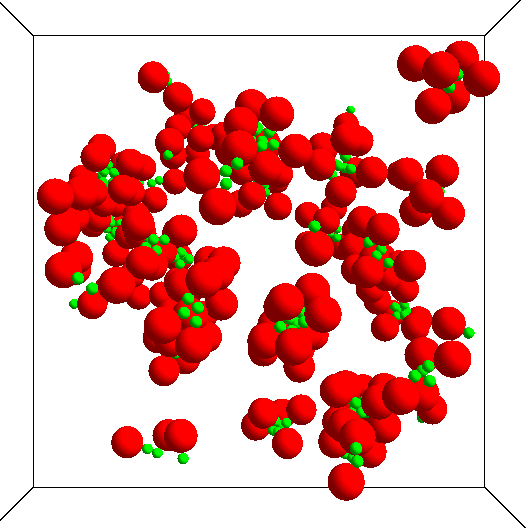}
\includegraphics[width=3.8cm,angle=0]{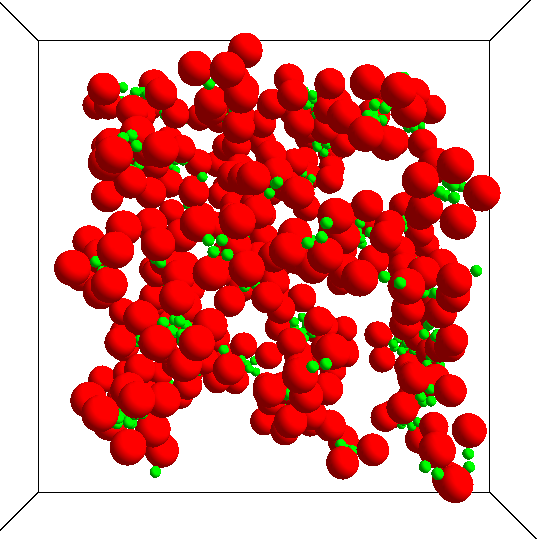}
\includegraphics[width=3.8cm,angle=0]{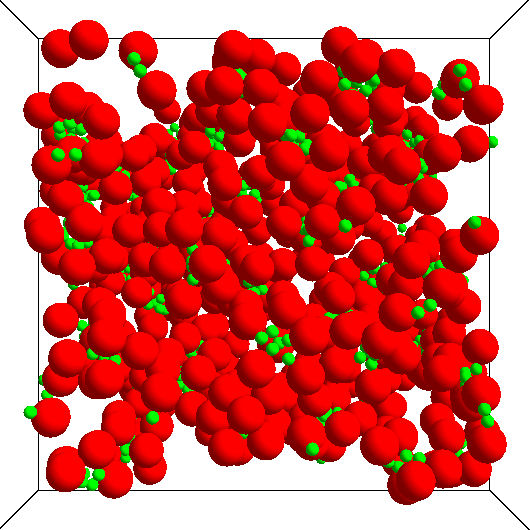}
\end{tabular}
\caption{Typical microscopic configurations
with $\alpha=0.33$ at $T^*=0.30$ (top panels)
and $T^*=0.20$ (bottom panels) and increasing densities
(from left to right, $\rho^*=0.05$, 0.10, 0.20 and 0.30).
}
\label{fig:snap-033}
\end{center}
\end{figure}

We recall that for $\alpha < 1$
the attractive (SW) sphere is smaller
than the repulsive (HS) one. We first consider $\alpha=0.33$.
The SW-SW structure factors $S_{11}(k)$
are shown in figure~\ref{fig:alfa-033}a
at $T^*=0.30$ (i.e.  slightly above $T^*_{\rm B}$ according to figure~\ref{fig:Boyle}),
and for various densities.
 All $S_{11}(k)$ show a low-$k$ peak, becoming
 more pronounced as the density increases, whereas
 $S_{11}(k\to 0)$ remain limited, such a behaviour  appearing
 compatible with the formation of a cluster fluid
 that suppresses the gas-liquid phase separation
(see also~\cite{Schurt:04}).
Indeed, both experiments~\cite{Schurt:04,liu:05}
and theoretical studies (see e.g.~\cite{Cardinaux:07,Costa:10,Bomont:12} and references)
point to such a low-$k$ peak as
indicating the formation of aggregates.
More generally, the presence of the low-$k$ peak has been recently
related to the onset of some kind of
``intermediate-range order'' in the 
fluid~\cite{liu:11,falus:12,Godfrin:14}. \\
The formation of the cluster fluid
can be further investigated by monitoring
he average internal energy per particle $\langle E\rangle$
(in units of $\varepsilon$) over 
an extended temperature
range encompassing $T^*_{\rm B}$:
as visible from figure~\ref{fig:alfa-033}b,
the almost flat behaviour of $\langle E\rangle$ vs $\rho^*$
at $T^*=0.50$ is replaced by a monotonic decay at
$T^*=0.30$, becoming flat again at $T^*=0.20$. These outcomes indicate that, at
the highest temperature,
clusters are not able to develop, since $\langle E\rangle \approx 0$
regardless of the density. At $T^*=0.30$, instead,
the attractive energy is strong enough to drive a cluster assembly process,
provided the density is high enough, as signalled by the strong
enhancement of the low-$k$ peak of $S_{11}(k)$, occurring only
for $\rho^* \geq 0.2$ (see figure~\ref{fig:alfa-033}a).
Finally, if the temperature
is further lowered down to $T^*=0.20$,
the system is able to assemble into clusters even at low density;
indeed, at this temperature,
the energy attains almost the same (significantly negative)
value, irrespective of the density.
Typical
microscopic configurations at $T^*=0.30$ and 0.20, displayed in
figure~\ref{fig:snap-033}, confirm the above picture;
note in particular that at $T^*=0.20$ (bottom panels) clusters
are visible even in the low density regime.
It is worth noting that, under appropriate conditions
for the development of clusters, the ratio $|\langle E\rangle|/T^*$ 
ranges approximately between 3 and 10, 
thus confirming the relative
stability of aggregates; on the other hand, this does 
not preclude the possibility for particles to rearrange 
within the clusters, as well as to be exchanged 
between clusters and the surrounding medium,
as also observed in experiments~\cite{Kraft:12}).
%
\begin{figure}[!t]
\begin{center}
\begin{tabular}{lr}
\includegraphics[width=8.0cm,angle=0]{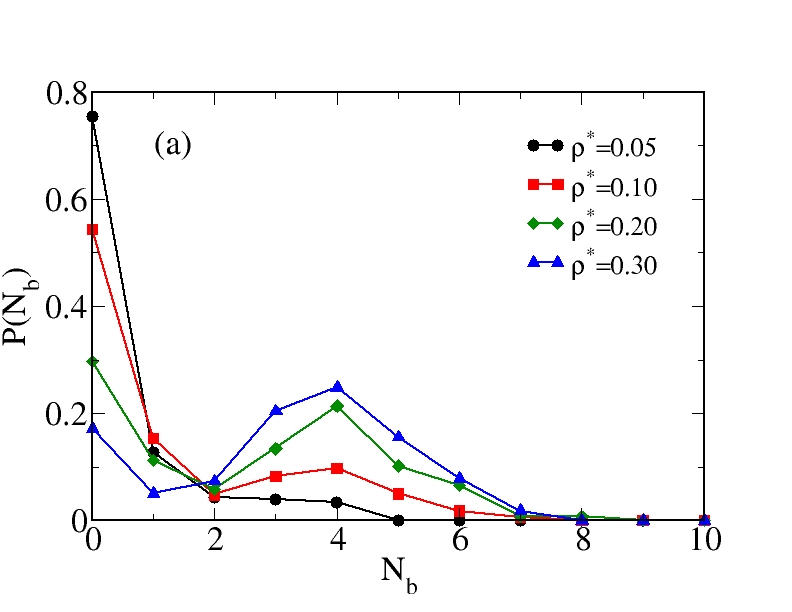} &
\includegraphics[width=8.0cm,angle=0]{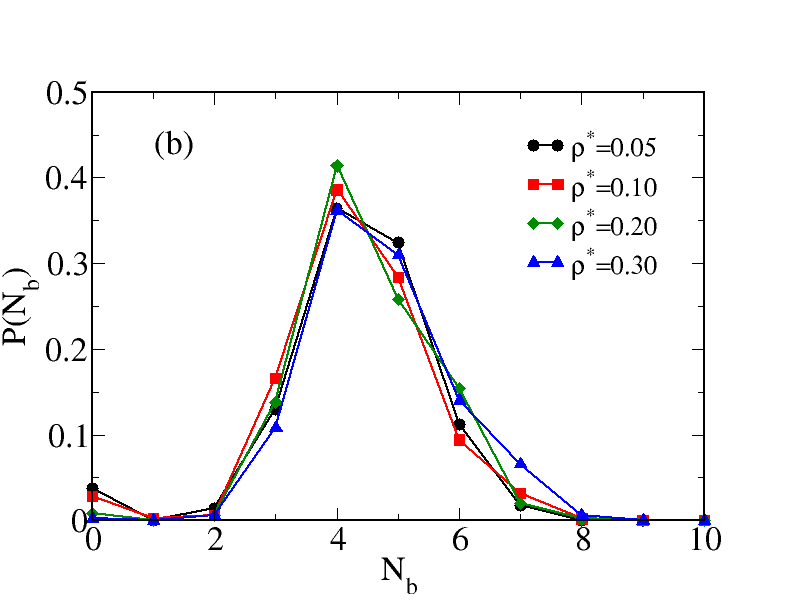}
\end{tabular}
\caption{Probability distribution of bonds
with $\alpha=0.33$ for several densities
at (a) $T^*=0.30$ and (b) $T^*=0.20$. Lines are guides to the eye.
}
\label{fig:bonds-033}
\end{center}
\end{figure}
Our analysis is further supported by the 
probability distribution of bonds,
$P(N_b)$  shown in figure~\ref{fig:bonds-033}~---~%
where $N_b$ is the number of bonds per particle, assuming two particles as bonded
together
if the distance between the SW spheres falls within the
corresponding attraction range, i.e. between $\sigma_1$ and $\sigma_1 +\lambda
\sigma_1$.
One can notice that
at $T^*=0.30$ (panel a), a maximum value is attained
for $N_b=0$ at $\rho^*=0.05$, this suggesting
that few bonds among HJD are formed in such conditions. The maximum
shifts towards higher values of $N_b$ only upon increasing $\rho^*$.
Things
drastically change when $T^*=0.20 < T^*_{\rm B}$ (panel b), with
the maximum of $P(N_b)$
centred around $N_b=4$ and almost insensitive to density variations.
Note that the value attained by the internal energy
$|\langle E\rangle| \approx2$ when $T^*=0.20$
(see
figure~\ref{fig:bonds-033}b)
is congruent
with a bond configuration in which each dumbbell forms four bonds,
since $|\langle E\rangle|$ scales with $N_b/2$.
We have observed that at $T^*=0.20$ and for the various 
densities examined in this work, clusters are formed by three to ten particles.
In such conditions, we have preliminarily calculated the average 
radius of gyration of clusters $\langle R_{\rm g} \rangle$,
as function of their size. 
We have found that specific values of $\langle R_{\rm g} \rangle$ attained 
for a given cluster size do not practically depend on the density of the 
system. In general, the trend observed for $\langle R_{\rm g} \rangle$  
as a function of the cluster size, along with visual inspection of cluster 
configurations (see especially bottom panels of figure 4), suggests that 
roughly spherical aggregates develop in the system.

\begin{figure}[b!]
\begin{center}
\begin{tabular}{lr}
\includegraphics[width=8.0cm,angle=0]{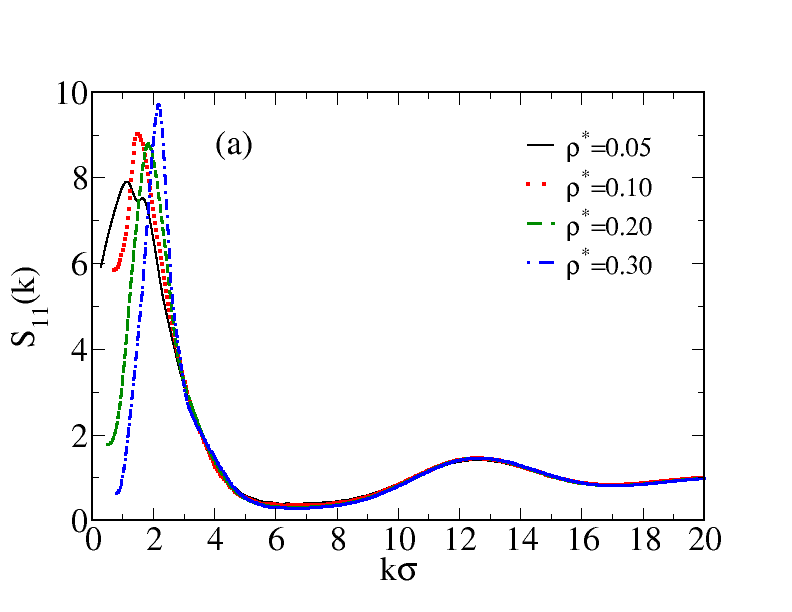} &
\includegraphics[width=8.0cm,angle=0]{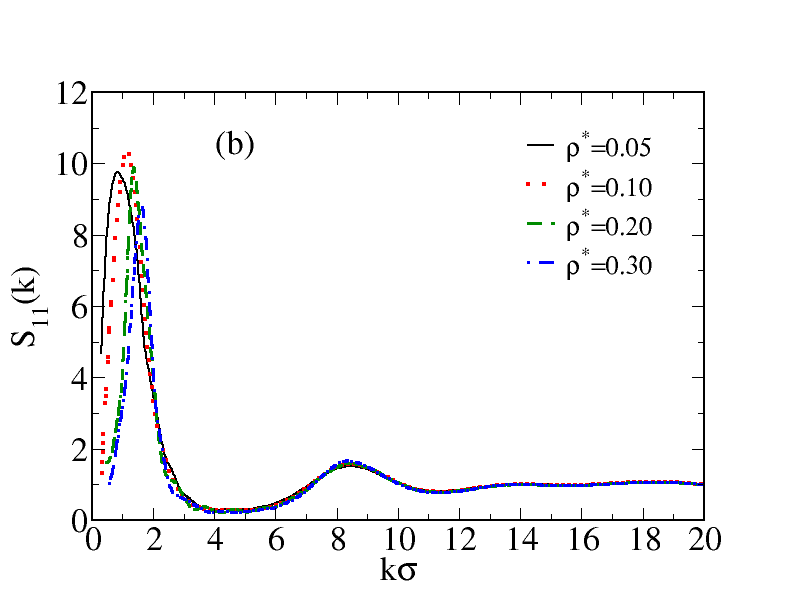}
\end{tabular}
\caption{$S_{11}(k)$ at $T^*=0.30$ and
several densities with (a) $\alpha=0.50$ and (b) $\alpha=0.75$.}
\label{fig:sk-050}
\end{center}
\end{figure}
\begin{figure}
\begin{center}
\begin{tabular}{lr}
\includegraphics[width=8.0cm,angle=0]{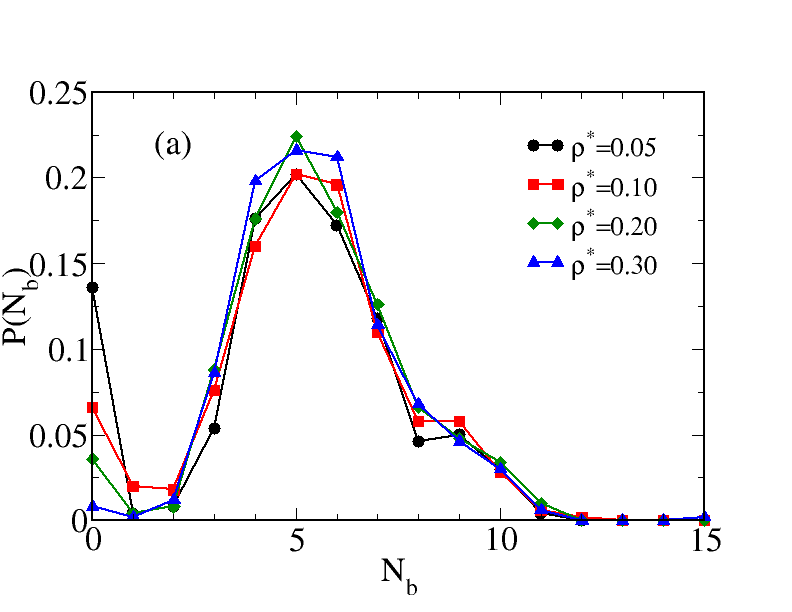} &
\includegraphics[width=8.0cm,angle=0]{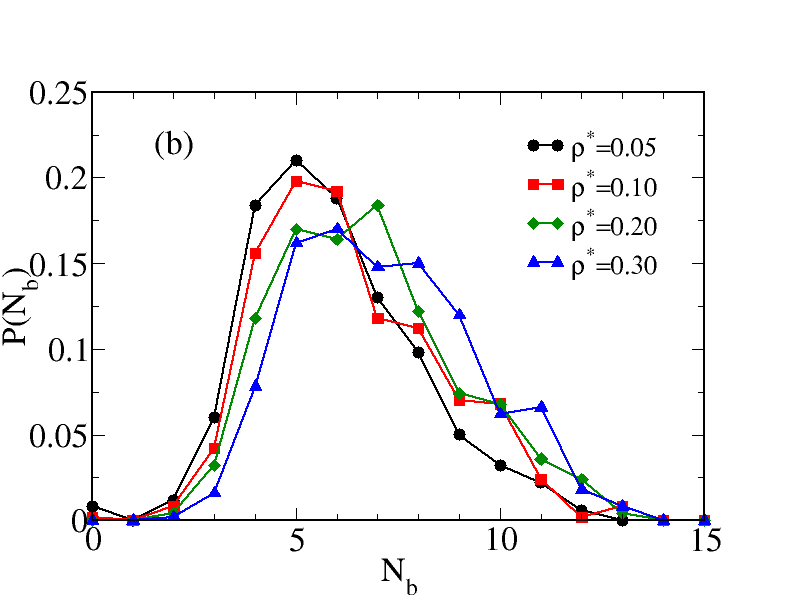}
\end{tabular}
\caption{Probability distribution of bonds at $T^*=0.30$ and several densities
with (a) $\alpha=0.50$ and (b) $\alpha=0.75$. Lines are guides to the eye.}
\label{fig:bonds-075}
\end{center}
\end{figure}
\begin{figure}
\begin{center}
\includegraphics[width=3.0cm,angle=0]{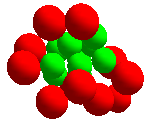}
\caption{Sketch of a cluster composed by ten HJD
molecules with $\alpha=0.75$ at $T^*=0.30$, as obtained by MC simulations.}
\label{fig:micelles}
\end{center}
\end{figure}
Upon increasing $\alpha$, but still keeping below $\alpha=1$,
$T^*_{\rm B}$ increases (see figure~\ref{fig:Boyle}) and HJD can bond together
already at higher $T^*$.
However, the structural properties
do not change qualitatively:
$S_{11}(k)$ still display a low-$k$ peak
whereas $S_{11}(k\to 0)$ remain limited, as shown in figure~\ref{fig:sk-050}
where results concerning $\alpha=0.5$ and $\alpha=0.75$ at
$T^*=0.30$ are reported.
Some new features emerge in
the probability distribution of bonds, as a consequence of the larger
number of HJD that can assemble into a cluster:
specifically, $P(N_b)$, reported in figure~\ref{fig:bonds-075},
now displays multiple peaks at high densities indicating
an increased inhomogeneity in the dumbbell bonding environments.
Such features appear compatible
with two distinct components of the clusters, namely capping dumbbells,
constituting either the total surface of the aggregate or caps of
elongated structures (associated to the first peak), and bulk dumbbells
(associated to the second peak). At low densities, 
when smaller clusters
are favoured, only the first configuration occurs.
When the average cluster
size becomes larger than $\approx 10$, both capping and bulk arrangements
are observed yielding an additional peak.
A more detailed analysis of particle distribution inside the cluster
seems in order to better resolve this possibility.

Interestingly enough, the above findings
qualitatively agree with experimental
results~\cite{Kraft:12} on self-assembly of dumbbell-shaped
particles in colloidal micelles. Here, results are presented
concerning
synthesized dumbbell particles with one rough (``hard-sphere'') and one smooth
(``square-well'') sphere, interacting through depletion interactions.
As the size ratio between smooth and rough sphere falls around
(1.11/1.46) micron, particles self-assemble into colloidal micelles,
a result in close correspondence with our observations in the
same conditions (i.e. with $\alpha \sim0.75$), as documented
for instance
by the snapshot taken from our MC simulation reported in figure~\ref{fig:micelles}.

\subsection{Competition between cluster formation and phase separation
(${\alpha=1.10}$)}

\begin{figure}
\begin{center}
\includegraphics[width=8.0cm,angle=0]{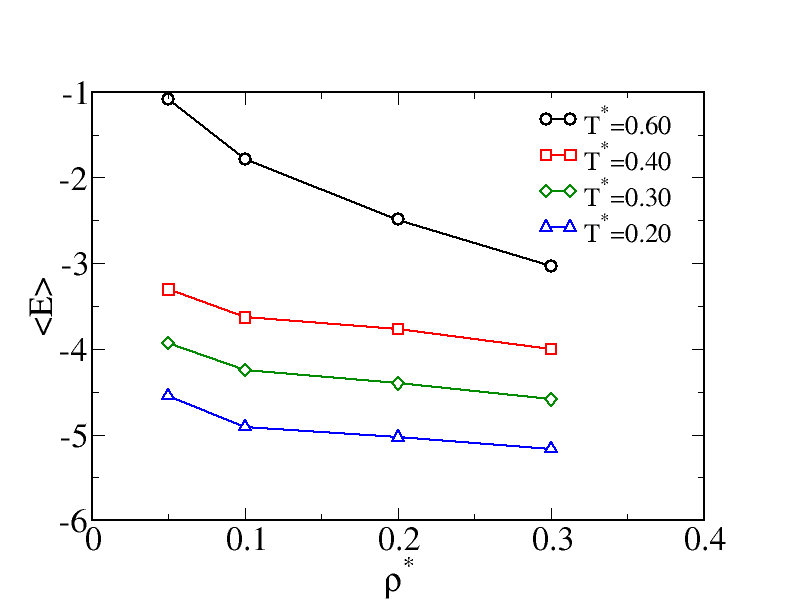}
\caption{Average internal energy per particle $\langle E \rangle$
with $\alpha=1.10$
as a function of the density along different isotherms.}\label{fig:energy}
\end{center}
\end{figure}

\begin{figure}[!t]
\begin{center}
\begin{tabular}{lr}
\includegraphics[width=8.0cm,angle=0]{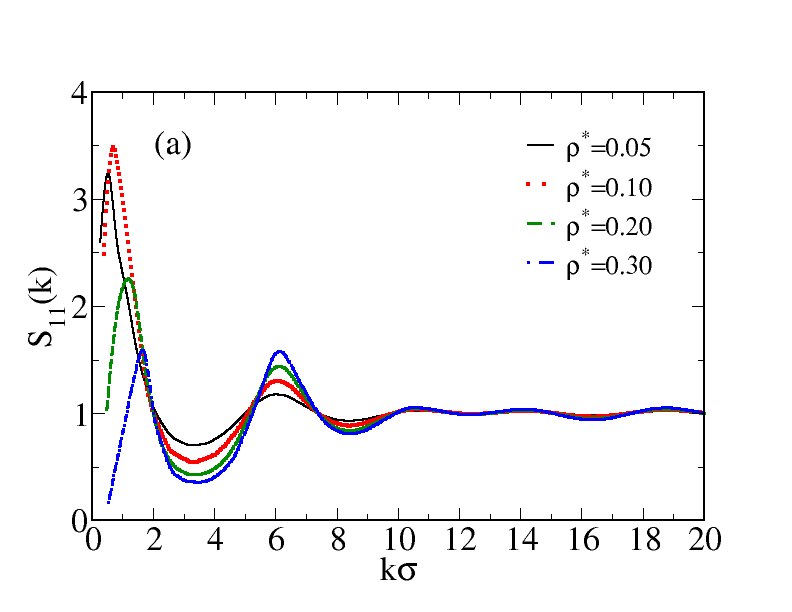} &
\includegraphics[width=8.0cm,angle=0]{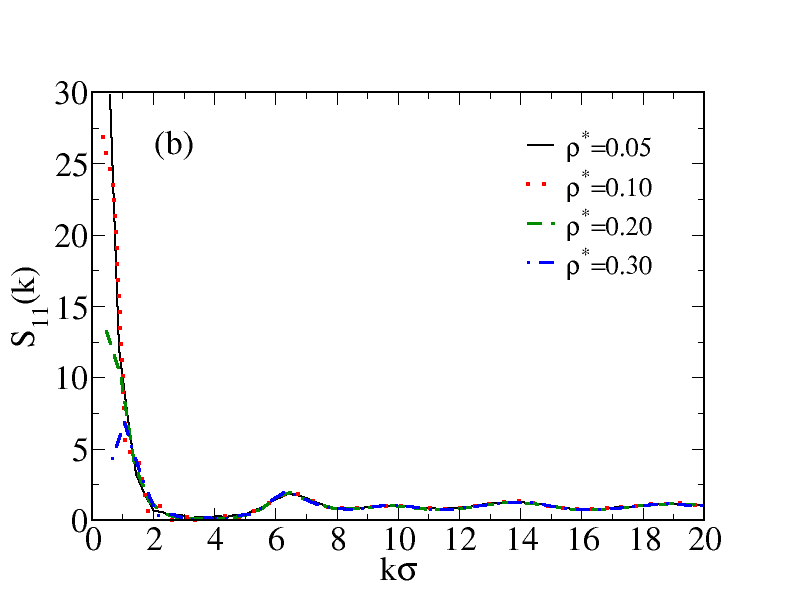}
\end{tabular}
\caption{$S_{11}(k)$
with $\alpha=1.10$ and for several densities at (a) $T^*=0.60$,
and (b) $T^*=0.20$.}\label{fig:sk}
\end{center}
\end{figure}

As we have shown in Ref.~\cite{Munao:14} no gas-liquid coexistence
takes place at $\alpha=1$, and results from Section~\ref{sec:3.1}
indicate that this holds for all $\alpha <1$.
Such a scenario changes remarkably as soon as
$\alpha$ gets larger than unity.
We first consider
results for $\langle E \rangle$ at $\alpha=1.10$,
reported in figure~\ref{fig:energy}: the monotonic decrease of the energy
with increasing density, visible at high temperatures, is progressively
replaced by an almost flat trend as the temperature goes down to
$T^*=0.2$, where the internal energy is almost constant for all $\rho^*$.
The absence of
jumps in $\langle E \rangle$ suggests that no large-scale
aggregates (as, for instance, lamell\ae) are formed in the system.
On the other hand, $S_{11}(k)$ at $T^*=0.60$
(shown in figure~\ref{fig:sk}a) clarly displays
a low-$k$ peak
for all densities, this feature
suggesting the development of clusters
in the system. Conversely, if the temperature is lowered to $T^*=0.20$
(figure~\ref{fig:sk}b),
the low-$k$ peak disappears
at low and intermediate densities and survives only at $\rho^*=0.30$;
the disappearance of the low-$k$ peak gives place to the
rise of the $k\to 0$ limit of $S_{11}(k)$,
suggesting that a phase separation process is taking place.

\begin{figure*}
\begin{center}
\begin{tabular}{l}
\includegraphics[width=3.7cm,angle=0]{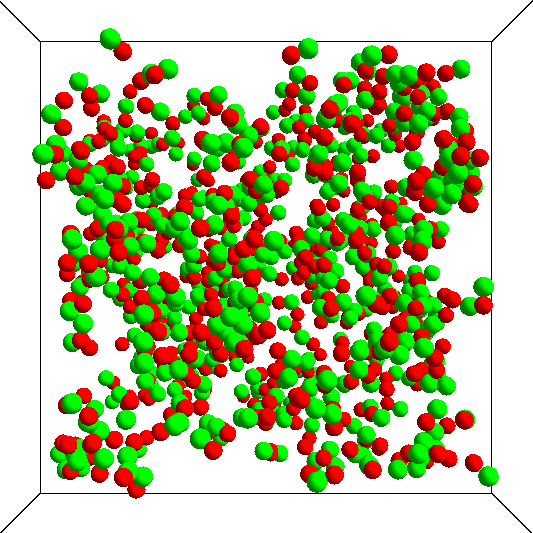}
\includegraphics[width=3.7cm,angle=0]{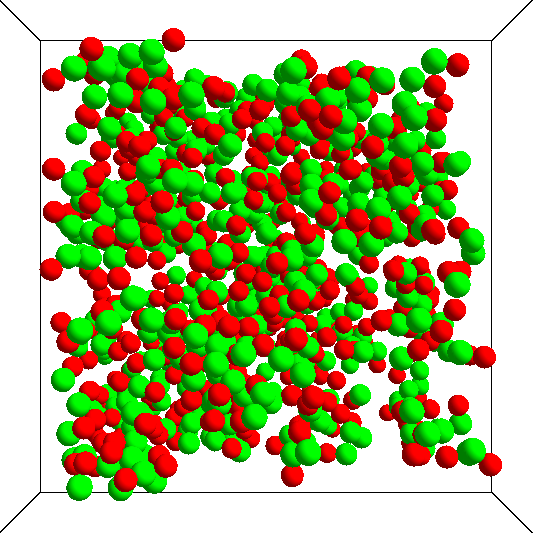}
\includegraphics[width=3.7cm,angle=0]{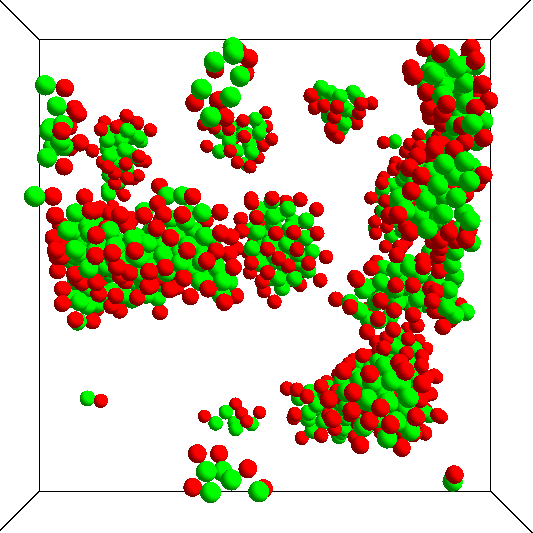}
\includegraphics[width=3.7cm,angle=0]{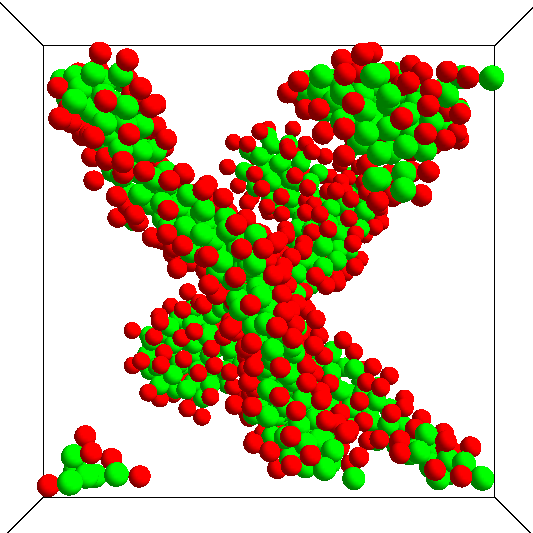} \\
\hspace{45pt} (a) \hspace{85pt} (b) \hspace{85pt} (c) \hspace{85pt} (d)
\end{tabular}
\caption{Typical microscopic configurations
with $\alpha=1.10$
at $T^*=0.60$ (a, b) and 0.20 (c, d) and
$\rho^*=0.05$ (a, c), and 0.10 (b, d).}
\label{fig:snap-110}
\end{center}
\end{figure*}

Such observations are compatible with a
competition between self-assembly processes (at high temperature)
and gas-liquid phase coexistence (at lower temperatures).
Note, in this instance, that here, unlike previous cases,
both $T^*=0.60$ and $T^*=0.20$
are lower than $T^*_{\rm B}$ (see figure~\ref{fig:Boyle}), this
indicating that the competition between the two regimes occur when the
attractive part of the interaction is significant.
Visual evidence is offered by
snapshots of microscopic configurations displayed in
figure~\ref{fig:snap-110}: HJD assemble in small
clusters at $T^*=0.60$ and densities $\rho^*=0.05$ and
$\rho^*=0.10$ (first two panels);
upon lowering the temperature, the phase separation process
dominates, as visible in the last two panels
corresponding to $T^*=0.20$ and $\rho^*=0.05$ and
$\rho^*=0.10$.

\subsection{Phase separation (${1.25\leq\alpha\leq1.66}$)}

In this regime the SW interaction becomes significantly larger
than the HS one. The ensuing reduced role of short-range repulsion
should favour the phase separation mechanism.
\begin{figure}[!h]
\begin{center}
\begin{tabular}{lr}
\includegraphics[width=8.0cm,angle=0]{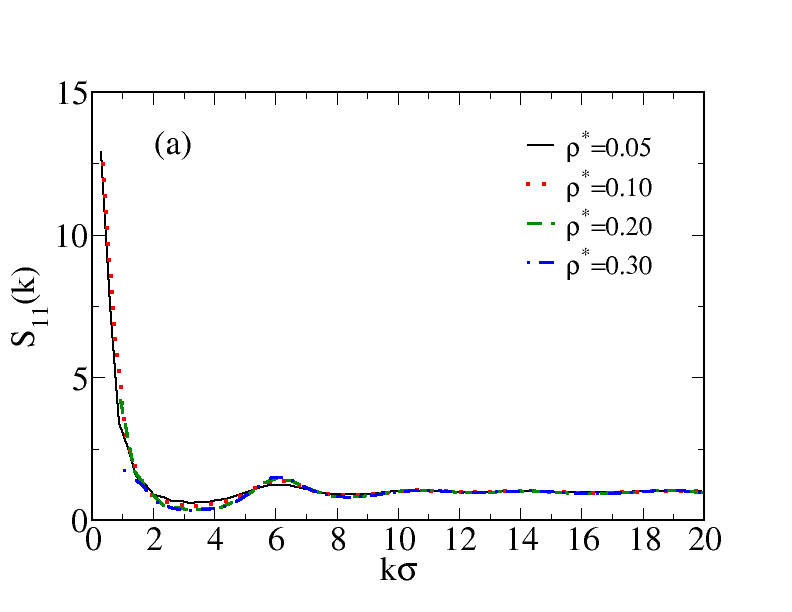} &
\includegraphics[width=8.0cm,angle=0]{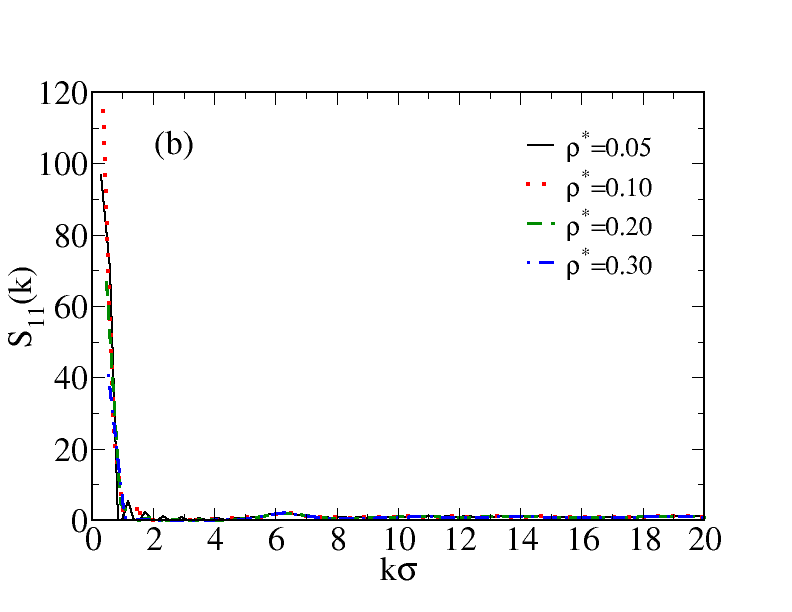}
\end{tabular}
\caption{$S_{11}(k)$ at $T^*=0.60$
and for several densities
with (a) $\alpha=1.25$ and (b) $\alpha=1.66$.}
\label{fig:sk-125}
\end{center}
\end{figure}
We illustrate this point through simulations carried out at
$\alpha=1.25$ and $\alpha=1.66$, for $T^*=0.60$
(i.e. below $T^*_{\rm B}$, see figure~\ref{fig:Boyle}) and
increasing densities. All $S_{11}(k)$, reported in figure~\ref{fig:sk-125},
show a clear diverging trend with $k\to 0$, thus indicating that the
system is close to (or has already
crossed) a metastable region. This observation is supported by snapshots
reported in figure~\ref{fig:snap-125}, where no clusters are observed,
either at high or low density; the phase-separation process is instead
clearly visible in panels (c) and (d), where the system appears
separated into gas and liquid regions.

\begin{figure}[!t]
\begin{center}
\begin{tabular}{l}
\includegraphics[width=3.7cm,angle=0]{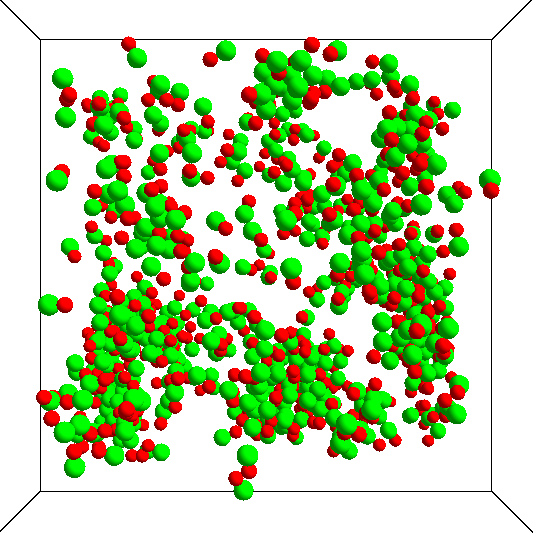}
\includegraphics[width=3.7cm,angle=0]{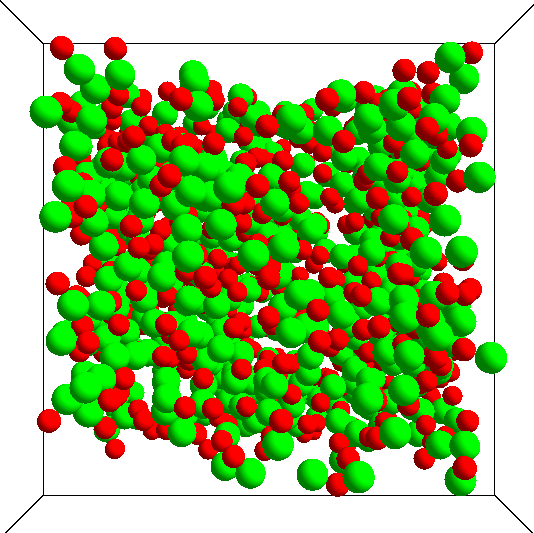}
\includegraphics[width=3.7cm,angle=0]{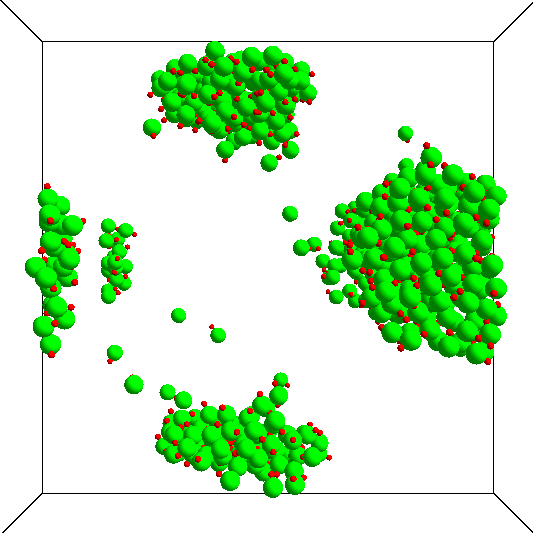}
\includegraphics[width=3.7cm,angle=0]{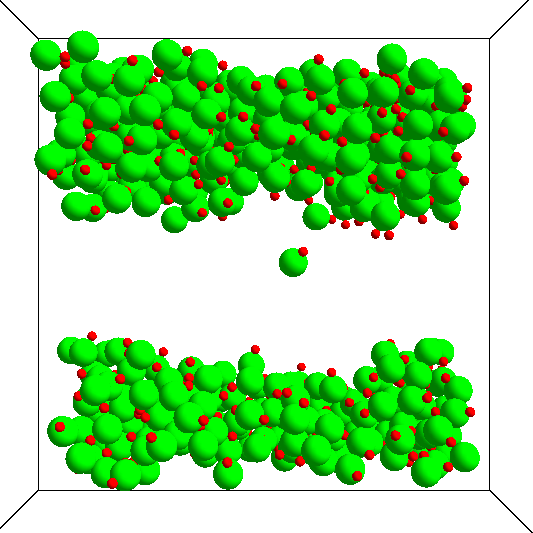} \\
\hspace{45pt} (a) \hspace{85pt} (b) \hspace{85pt} (c) \hspace{85pt} (d)
\end{tabular}
\caption{Typical microscopic configurations at $T^*=0.60$
with $\alpha=1.25$ (a, b) and $\alpha=1.66$ (c, d)
and $\rho^*=0.05$ (a, c) and $\rho^*=0.20$ (b, d).}
\label{fig:snap-125}
\end{center}
\end{figure}

\begin{figure}
 \begin{center}
 \includegraphics[width=15.0cm,angle=0]{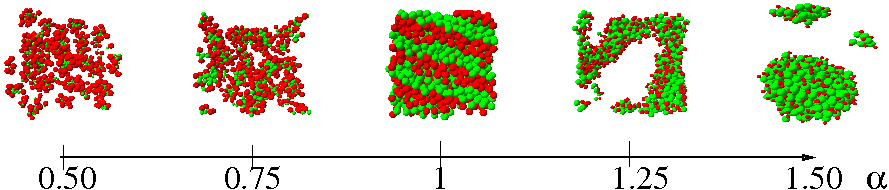}
 \caption{Phase behaviour of HJD
 as a function of $\alpha$ (at fixed $T^*=0.30$ and $\rho^*=0.10$).
 As visible, the planar configuration (lamell\ae)
 is observed only around the homonuclear case ($\alpha=1$).}
 \label{fig:lamellae}
 \end{center}
 \end{figure}

 \begin{table}[!h]
 \caption{Phase behaviour of HJD as a
 function of $\alpha$. Corresponding values
of the HS ($\sigma_1$) and SW ($\sigma_2$) diameters
are indicated. Note that if $\alpha=0$ the model reduces
to a single hard-sphere particle and, at the opposite end, to a single
square-well particle.  }
 \label{tab:phase}
 \begin{center}
 \begin{tabular}{cccccc}
 \hline\hline
 $\alpha$ & $\sigma_1$ & $\sigma_2$ & Clusters & Lamellae & Phase Separation \\
 \hline
 0.00 & 0.00 & 1.00 & \xmark & \xmark & \xmark \\
 0.33 & 0.33 & 1.00 & \cmark & \xmark & \xmark \\
 0.50 & 0.50 & 1.00 & \cmark & \xmark & \xmark \\
 0.75 & 0.75 & 1.00 & \cmark & \xmark & \xmark \\
 1.00 & 1.00 & 1.00 & \cmark & \cmark & \xmark \\
 1.10 & 1.00 & 0.90 & \cmark & \xmark & \cmark \\
 1.25 & 1.00 & 0.75 & \xmark & \xmark & \cmark \\
 1.50 & 1.00 & 0.50 & \xmark & \xmark & \cmark \\
 1.66 & 1.00 & 0.34 & \xmark & \xmark & \cmark \\
 2.00 & 1.00 & 0.00 & \xmark & \xmark & \cmark \\
 \hline\hline
 \end{tabular}
 \end{center}
 \end{table}
The
generic phase behaviour of HJD is schematically reported
in table~\ref{tab:phase}, where the presence (or absence) 
of clusters, lamell{\ae} and phase separation is recorded 
as a function of $\alpha$.
Such different arrangements  are also displayed in 
figure~\ref{fig:lamellae}, concerning specifically
the thermodynamic condition $T^*=0.30$ and $\rho^*=0.10$.
To summarize,
the competition between cluster formation and phase separation
favours the
former at low/intermediate values of $\alpha$ and the latter at
intermediate/high values of $\alpha$. Accordingly,
one can reasonably
surmise that the presence of both is
possible only over an intermediate narrow interval, namely for $\alpha$
between 1.00 and 1.10.
As a consequence,
a subtle equilibrium exists between phase separation and
self-assembly, strongly depending upon
the heterogeneity of HJD.
Lamellar structures are hardly
observed upon varying $\alpha$, this suggesting that
only $\alpha\approx 1$ is compatible with
the development of lamell{\ae}, as found in our previous work~\cite{Munao:14}.
There, the simultaneous presence of both clusters and lamell{\ae} has been
documented by the behaviour of an order parameter
quantifying the average relative orientations of dimers, and the rotational 
invariant of local bond 
order parameters $q_6$.
It is worth noting that
it does not exist a value of $\alpha$ compatible with the simultaneous
presence of clusters, lamell{\ae} 
and phase separation;
instead, as documented in~\cite{Munao:14}, this may happen
in specific cases for homonuclear square-well dumbbells.

\section{Conclusions}

We have investigated the self-assembly process and
gas-liquid phase separation taking place in heteronuclear Janus dumbbells
(HJD), modelled by two tangent hard spheres with different
core diameters, the first one being
surrounded by a square-well attraction.
We have carried out
standard Monte Carlo simulations
to characterize the fluid structure, the distribution of bonds
among molecules,
and to study the formation of clusters and phase separations.

The relative size of the two spheres constituting one HJD molecule
has been changed by introducing
a parameter $\alpha$:
for $\alpha <1$ (corresponding to the square-well
site smaller than the hard-sphere one), we have observed the development of
a cluster phase,
with spherical aggregates becoming increasingly
structured upon lowering the temperature. Here,
no indication emerges on the presence of a gas-liquid critical behaviour,
thus suggesting that the cluster formation suppresses the phase separation.
These findings qualitatively agree with experimental
results cataloguing the self-assembly of 
dumbbell-shaped particles~\cite{Kraft:12}.
Moving towards $\alpha=1$, the square-well attraction increases,
allowing for a progressively large number of bonds per molecule
to be established; as a consequence,
HJD may self-assemble into relatively large
clusters of different sizes and shapes. 
In the case $\alpha=1$, corresponding
to homonuclear Janus dumbbells, we
have previously documented~\cite{Munao:14}
the absence of a gas-liquid coexistence
and the simultaneous appearance of
planar structures (lamell\ae). The development of these
latter turns to depend sensitively on the symmetry of
dumbbells, and therefore is essentially confined to the homonuclear
case.
Finally, if $\alpha$ is further increased, the attractive interaction
becomes more and more isotropic and the
gas-liquid phase separation progressively
dominates, first competing with (till $\alpha=1.10$),
then completely suppressing the
formation of clusters.

Our model may constitute a useful prototype to investigate the
role of size asymmetry and attractive interactions
in the phase behaviour of dumbbell-shaped colloids
with different chemical compositions, allowing for a deeper understanding
of the competition between self-assembly  and
phase separation in such systems.

\section*{Acknowledgements}
GM, DC, CC, FS and AG gratefully acknowledge support from PRIN-MIUR
2010-2011 project. AG, PO'T and TSH acknowledge support of a 
Cooperlink bilateral agreement Italy-Australia.

\providecommand{\newblock}{}

\end{document}